# Surface air temperature variability in global climate models

Richard Davy and Igor Esau
G.C. Rieber Climate Institute of the Nansen Environmental and Remote Sensing Centre, Bjerknes Centre for Climate Research, Thormøhlensgt. 47, 5006, Bergen, Norway.

Correspondence to: richard.davy@nersc.no

New results from the Coupled Model Inter-comparison Project phase 5 (CMIP5) and multiple global reanalysis datasets are used to investigate the relationship between the mean and standard deviation in the surface air temperature. A combination of a land-sea mask and orographic filter were used to investigate the geographic region with the strongest correlation and in all cases this was found to be for low-lying over-land locations. This result is consistent with the expectation that differences in the effective heat capacity of the atmosphere are an important factor in determining the surface air temperature response to forcing.

## 1. Introduction

The surface air temperature (SAT) and its various properties (diurnal and seasonal range and extremes, trends and variability) are some of the most widely used metrics of global climate and climate change [IPCC AR4, 2007]. This is to be expected given SAT is such a readily measurable, and anthropically important, variable. The variability of the SAT at a given timescale is determined by three factors: the magnitude of the forcing, any feedback processes and the effective heat capacity of the system [Hasselmann, 1976]. Given the inverse dependency of the magnitude of changes in SAT on the effective heat capacity, some correlations between metrics of the SAT are anticipated to be innate to the thermodynamics of the climate system. For example, there is the strong positive correlation between SAT trends and variability across the globe, as has been shown in observations, reanalysis and CMIP3 datasets [Esau et al., 2012]. This is to be expected given that SAT trends and variability are both manifestations of the magnitude of SAT response to forcing.

The effective heat capacity of the atmosphere has been shown to be directly proportional to the depth of the planetary boundary layer (PBL) [Esau and Zilitinkevich, 2010]. The shallowest boundary-layers are found over land in locations with consistent negative surface sensible heat flux, leading to strongly stable thermal stratification. We can expect to see the strongest signal of PBL response at over-land locations. Since this mechanism is driven by the turbulent mixing that characterises the PBL we expect there to be a weaker relation between the temperature mean and variability in high-altitude locations where the PBL is exposed to the free atmosphere. There has been some indication that such patterns, predicted from PBL response theory, are constrained to the PBL from the work of Weber et al. [1994] but the reason for this was not identified. The prediction that this relationship between the mean and SD of SAT should be weaker in locations exposed to the free atmosphere allows us to falsify the PBL-response mechanism by comparing correlations of mean and SD of SAT with and without the use of an orographic filter.

It has been established that the coldest locations have the greatest SAT variability [IPCC AR4, 2007], but a more general relationship between the mean and variability of SAT can be masked by specific geographical conditions. Here we have used a simple least-squares linear regression analysis to investigate the relationship between mean and standard deviation of SAT from CMIP5 results and various reanalysis datasets.

## 2. Data sets

The data used for this study were taken from the CMIP5 results and various global re-analysis projects, the data for which is hosted by the National Center for Atmospheric Research (NCAR). The temperature data were taken from the last 30 years of the 'historical' runs of the CMIP5, and from the same years in the reanalysis datasets with the exception of the Japanese 25-year reanalysis project (JRA25) which is limited to 25 years.

The historical simulations of the CMIP5 are run using changing conditions consistent with observations including: atmospheric composition, including anthropogenic and volcanic influences; solar forcing; land use changes and emissions or concentrations of short-lived species and natural and anthropogenic aerosols [Taylor *et al.*, 2009]. These simulations are performed by coupled ocean-atmosphere global climate models. We have used results from 8 of the groups that contributed to CMIP5: CSIRO Australia's CSIRO Mk3.5 [Gordon *et al.*, 2010]; the Japan agency for Marine-Earth Science and Technology and National institute for environmental studies MIROC-ESM-CHEM [Watanabe *et al.*, 2011] and MIROC5 [Watanabe *et al.*, 2010]; the French national centre for meteorological research's CNRM-CM5 [Voldoire *et al.*, 2012]; the UK Met Office Hadley Centre's HadCM3 and HadGEM2-AO [Collins *et al.*, 2011]; the institute for numerical mathematics' INMCM4 [Volodin, et al., 2010] and the Norwegian Climate Centre's NorESM [Iversen *et al.*, 2012].

The reanalysis datasets used were: the JRA25; the National Centers for Environmental Prediction (NCEP) Climate Forecast System Reanalysis (CFSR); the European Center for Medium-range Weather Forecast (ECMWF) Interim reanalysis (ERA-Interim); the National Oceanic and Atmospheric Administration (NOAA) $20^{th}$ century reanalysis v2 and the NCEP/NCAR reanalysis I.

## 3. Methods

Each of the global climate models and reanalysis has associated orographic data. An orographic filter, 'Orog', was created that removed all data corresponding to locations where the surface elevation is greater than 1 km. Locations with surface elevation greater than 1km ('highland') are assumed to be exposed to the free atmosphere. The land-sea mask of each dataset was used as a 'land' filter, to select only those grid-points which lay over land. The combination of the orographic and land filters was used to select only low-lying over-land ('lowland') locations, where we anticipate the strongest signal of the PBL-response.

The monthly surface air temperatures were obtained for each dataset. A time-series of temperature anomalies was calculated by removing the all-time monthly-means from the temperature time series. The temperature statistics were calculated from the anomalies after the removal of a 4$^{th}$ order polynomial trend from the series [Braganza *et al*., 2003].

## 4. Results

Figure 1a shows the correlation of the mean and SD of the SAT from various global climate models. Application of the orographic filter always improves the correlation and the best correlations are found in lowland locations with the exception of the results from NorESM where the over-land correlation is especially poor. We find the same result in the reanalysis datasets with the strongest correlations always found for lowland locations (Figure 1b).

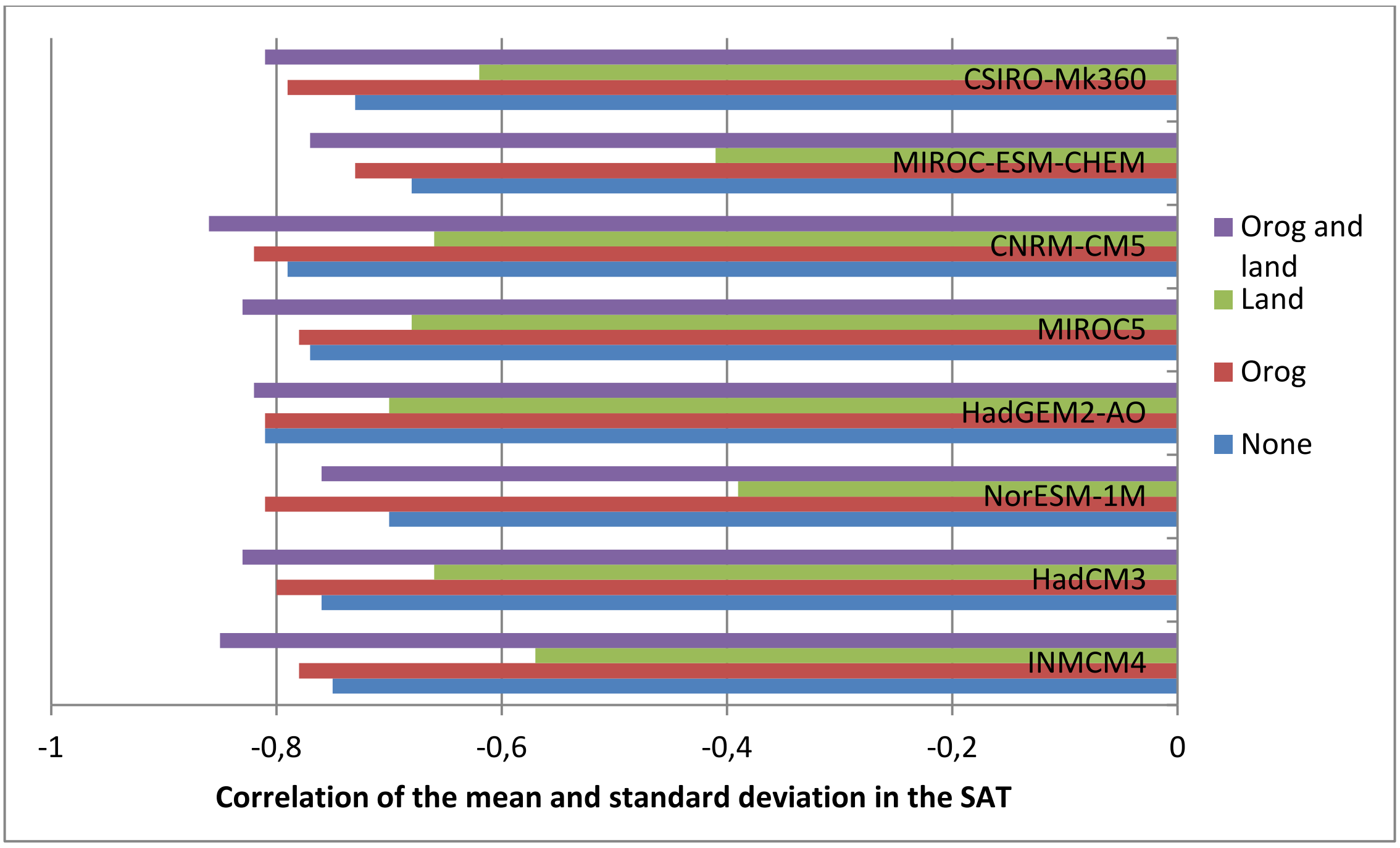

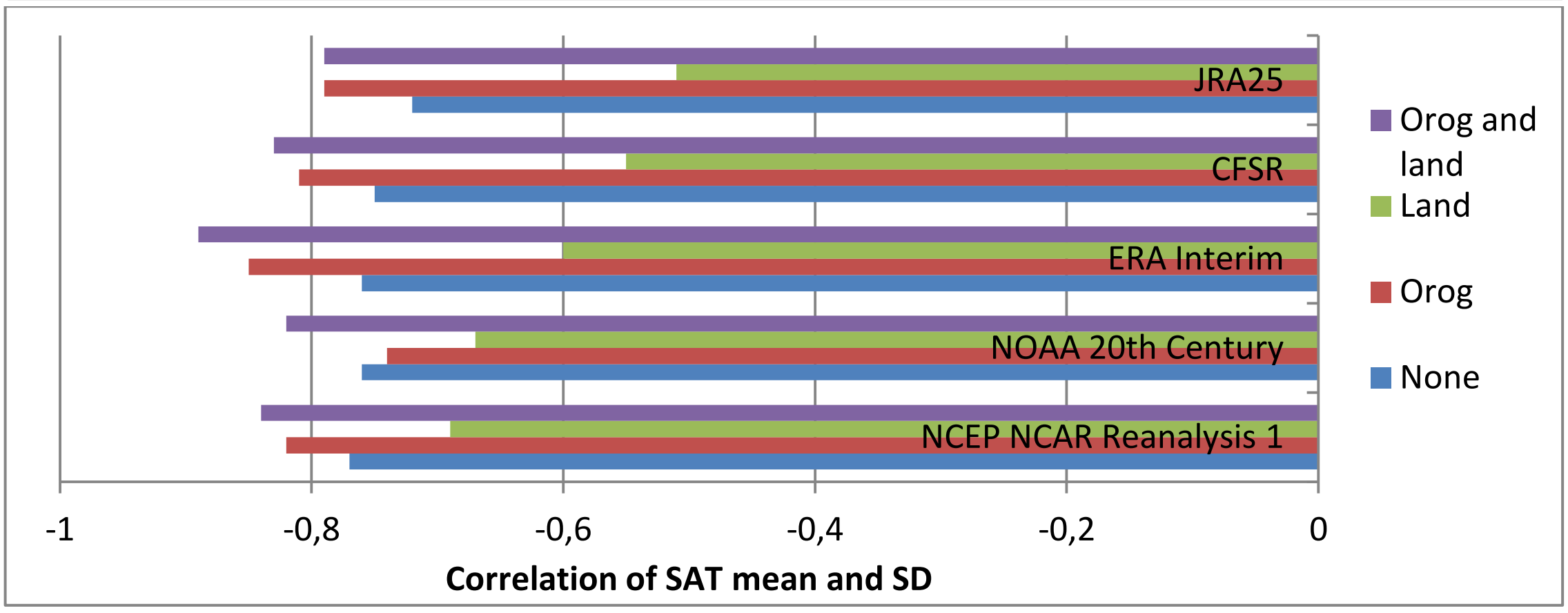

**Figure 1.** The correlation of the mean and standard deviation in surface air temperature from (a) CMIP5 models and (b) reanalysis calculated after applying a land-sea mask (land) and an orographic filter (Orog).

In all cases the biggest difference in the correlation comes with the addition of the orographic filter to the land-filtered data. The significance of these differences in the correlations can be seen from the difference in the number of data points used in each case (Figure 2). The highland regions represent 13% of the total number of grid points so the addition of the orographic filter to the global data does not significantly improve the correlation. However, the highland locations represent 41% of the over-land grid points; hence the largest improvement in the correlations is seen in the application of the orographic filter to the over-land data. This is also why we see a drop in the correlation after the application of the land filter to the global data: the fraction of grid points which correspond to highland locations is increased.

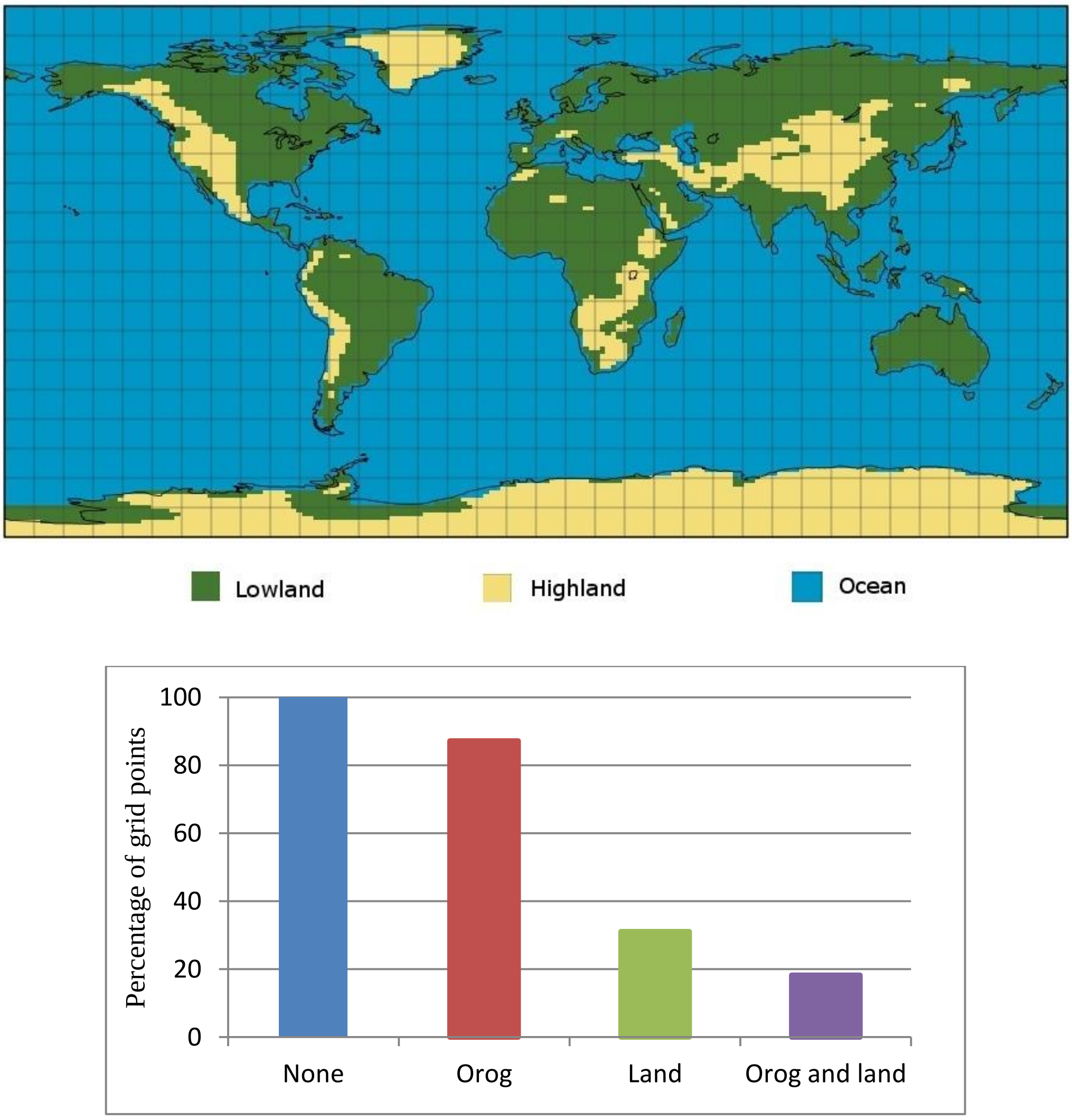

**Figure 2.** a) Global distribution of regions of ocean, high land and low-lying land. b) The percentage of grid points remaining after applying the land and/or orographic filters.

## 5. Conclusions

In this study we used the results from the latest phase of the CMIP to identify one of the climatological signatures of SAT variability: the strong negative correlation between SAT mean and variability. This result was confirmed through analysis of multiple reanalysis datasets. This relation is consistent with the PBL-response hypothesis: that differences in the SAT variability are, in part, due to variations in the effective heat capacity of the atmosphere such that the lower the heat capacity the greater the SAT response to a given forcing [Esau et al., 2012]. When we include the highland locations where the PBL is exposed to the free atmosphere, and as such is not expected to produce the same response, we find a weaker correlation between the SAT mean and SD. This is in agreement with our expectation that this relationship is a consequence of the PBL response.

**Acknowledgments.** This work has been funded by the Norwegian Research Council FRINAT project *PBL-feedback* 191516/V30.

## References

Balling Jr., R. C., P. J. Michaels and P. C. Knappenberger, (1998), Analysis of winter and summer warming rates in gridded temperature time series. *Clim. Res.*, **9**, 175-181, doi: 10.3354/cr009175.

Beare, R. J., et al., (2006), An intercomparison of Large-Eddy Simulations of the stable boundary layer. *Bound.-Lay. Meteorol.*, **118**, 247-272, doi: 10.1007/s10546-004-2820-6

Braganza, K., et al., (2003), Simple indices of global climate variability and change: Part I, Variability and correlation structure. *Clim. Dyn.*, **20**, 491-502.

Collins, W.J., et al., (2011), Development and evaluation of an Earth-System model – HadGEM2. *Geosci. Model Dev.*, **4**, 1051-1075, doi: 10.5194/gmd-4-1051-2011.

Cuxart, J., et al., (2006), Single-column model intercomparison for a stably stratified atmospheric boundary layer, *Bound.-Lay. Meteorol.*, **118**, 273-303, doi: 10.1007/s10546-005-3780-1

Easterling, D. R., et al., (1997), Maximum and minimum temperature trends for the globe, *Science*, **277**, 364-367.

Esau, I. and S. Zilitinkevich, (2010), On the role of the planetary boundary layer depth in the climate system. *Adv. Sci. Res.*, **4**, 63-69, doi:10.5194/asr-4-63-2010.

Esau, I., R. Davy and S. Outten, (2012), Complementary explanation of temperature response in the lower atmosphere. Submited to *Env. Res. Lett.* Aug 2012.

Gordon, H.B., et al., (2010), The CSIRO Mk3.5 Climate Model, technical report No. 21, The Centre for Australian Weather and Climate Research, Aspendale, Vic., Australia.

Hansen, J., M. Sato and R. Ruedy, (1995), Long-term changes of the diurnal temperature cycle: implications about mechanisms of global climate change. *Atmos. Res.*, **37**, 175-209, doi: 10.1016/0169-8095(94)00077-Q.

IPCC, (2007), Climate Change 2007: The Physical Science Basis. Contribution of Working Group I to the Fourth Assessment Report of the Intergovernmental Panel on Climate Change [Solomon, S., D. Qin, M. Manning, Z. Chen, M. Marquis, K.B. Averyt,

M.Tignor and H.L. Miller (eds.)]. Cambridge University Press, Cambridge, United Kingdom and New York, NY, USA.

Iversen, T., et al., (2012), The Norwegian Earth System Model, NorESM1-M – Part 2: Climate response and scenario projections. *Geosci. Model Dev. Discuss.*, **5**, 2933-2998.

Michaels, P. J., J. C. Balling Jr, R. S. Vose, P.C. Knappenberger, (1998), Analysis of trends in the variability of daily and monthly historical temperature measurements. *Clim. Res.* **10**, 27-33, doi: 10.3354/cr010027.

Taylor, K. E., R. J. Stouffer and G. A. Meehl, (2009), An overview of CMIP5 and the experiment design.*Bull.Amer. Meteor. Soc.*, **93**, 485-498.

Voldoire, A., et al., (2012), The CNRM-CM5.1 golbal climate model: description and basic Evaluation. *Clim. Dyn.*, doi: 10.1007/s00382-011-1259-y.

Volodin, E.M., N.A. Dianskii, A.V. Gusev, (2010), Simulating present-day climate with the INMCM4.0 coupled model of the atmospheric and oceanic general circulations. *Izvestiya Atmospheric and Oceanic physics*, **46** (4), 448-466.

Vose, R. S., D. R. Easterling, B. Gleason, (2005), Maximum and minimum temperature trends for the globe: An update through 2004. *Geophys. Res. Lett.* **32**, L23822.

Watanabe, S., et al., (2010), Improved climate simulation by MIROC5: Mean states, variability and climate sensitivity. *J. Climate*, **23**, 6312-6335, doi: 10.1175/2010JCLI3679.1

Watanabe, S., et al., (2011), MIROC-ESM 2012: model description and basic results of CMIP5-20c3m experiments. *Geosci. Model Dev.*, **4**, 845-872, doi: 10.5194/gmd-4-845-2011.

Weber, R. O., P. Talkner, G. Stefanicki, (1994), Asymmetric diurnal temperature change in the Alpine region. *Geophys. Res. Lett.* **21**, 673-676.

Zilitinkevich, S. S., et al., (2008), Turbulence energetics in stably stratified geophysical flows: strong and weak mixing regimes. *Quart. J. Roy. Met. Soc.*, **134**, 793-799, doi: 10.1002/qj.264.